\documentclass[10pt]{article}
\usepackage{spconf,amsmath,graphicx,enumitem}

\usepackage{cite}
\usepackage[T1]{fontenc}
\usepackage{graphicx}
\usepackage{amssymb}
\usepackage{amsmath}
\usepackage{paralist}
\usepackage{setspace}
\usepackage{microtype}
\usepackage{hyperref}
\usepackage{url}
\usepackage{balance}
\usepackage{xcolor}
\usepackage{fancyref}
\usepackage{dsfont}
\usepackage[font=small]{caption}

\usepackage{halloweenmath}

\usepackage{amssymb}
\usepackage{amsfonts}
\usepackage{mathrsfs}
\usepackage{xspace}
\usepackage{bm}
\usepackage{upgreek}

\newcommand{\safemath}[2]{\newcommand{#1}{\ensuremath{#2}\xspace}}

\safemath{\bma}{\mathbf{a}}
\safemath{\bmb}{\mathbf{b}}
\safemath{\bmc}{\mathbf{c}}
\safemath{\bmd}{\mathbf{d}}
\safemath{\bme}{\mathbf{e}}
\safemath{\bmf}{\mathbf{f}}
\safemath{\bmg}{\mathbf{g}}
\safemath{\bmh}{\mathbf{h}}
\safemath{\bmi}{\mathbf{i}}
\safemath{\bmj}{\mathbf{j}}
\safemath{\bmk}{\mathbf{k}}
\safemath{\bml}{\mathbf{l}}
\safemath{\bmm}{\mathbf{m}}
\safemath{\bmn}{\mathbf{n}}
\safemath{\bmo}{\mathbf{o}}
\safemath{\bmp}{\mathbf{p}}
\safemath{\bmq}{\mathbf{q}}
\safemath{\bmr}{\mathbf{r}}
\safemath{\bms}{\mathbf{s}}
\safemath{\bmt}{\mathbf{t}}
\safemath{\bmu}{\mathbf{u}}
\safemath{\bmv}{\mathbf{v}}
\safemath{\bmw}{\mathbf{w}}
\safemath{\bmx}{\mathbf{x}}
\safemath{\bmy}{\mathbf{y}}
\safemath{\bmz}{\mathbf{z}}
\safemath{\bmzero}{\mathbf{0}}
\safemath{\bmone}{\mathbf{1}}

\bmdefine{\biad}{a}
\bmdefine{\bibd}{b}
\bmdefine{\bicd}{c}
\bmdefine{\bidd}{d}
\bmdefine{\bied}{e}
\bmdefine{\bifd}{f}
\bmdefine{\bigd}{g}
\bmdefine{\bihd}{h}
\bmdefine{\biid}{i}
\bmdefine{\bijd}{j}
\bmdefine{\bikd}{k}
\bmdefine{\bild}{l}
\bmdefine{\bimd}{m}
\bmdefine{\bind}{n}
\bmdefine{\biod}{o}
\bmdefine{\bipd}{p}
\bmdefine{\biqd}{q}
\bmdefine{\bird}{r}
\bmdefine{\bisd}{s}
\bmdefine{\bitd}{t}
\bmdefine{\biud}{u}
\bmdefine{\bivd}{v}
\bmdefine{\biwd}{w}
\bmdefine{\bixd}{x}
\bmdefine{\biyd}{y}
\bmdefine{\bizd}{z}

\bmdefine{\bixid}{\xi}
\bmdefine{\bilambdad}{\lambda}
\bmdefine{\bimud}{\mu}
\bmdefine{\bithetad}{\theta}
\bmdefine{\biphid}{\phi}
\bmdefine{\bideltad}{\delta}

\safemath{\bmia}{\biad}
\safemath{\bmib}{\bibd}
\safemath{\bmic}{\bicd}
\safemath{\bmid}{\bidd}
\safemath{\bmie}{\bied}
\safemath{\bmif}{\bifd}
\safemath{\bmig}{\bigd}
\safemath{\bmih}{\bihd}
\safemath{\bmii}{\biid}
\safemath{\bmij}{\bijd}
\safemath{\bmik}{\bikd}
\safemath{\bmil}{\bild}
\safemath{\bmim}{\bimd}
\safemath{\bmin}{\bind}
\safemath{\bmio}{\biod}
\safemath{\bmip}{\bipd}
\safemath{\bmiq}{\biqd}
\safemath{\bmir}{\bird}
\safemath{\bmis}{\bisd}
\safemath{\bmit}{\bitd}
\safemath{\bmiu}{\biud}
\safemath{\bmiv}{\bivd}
\safemath{\bmiw}{\biwd}
\safemath{\bmix}{\bixd}
\safemath{\bmiy}{\biyd}
\safemath{\bmiz}{\bizd}

\safemath{\bmxi}{\bixid}
\safemath{\bmlambda}{\bilambdad}
\safemath{\bmmu}{\bimud}
\safemath{\bmtheta}{\bithetad}
\safemath{\bmphi}{\biphid}
\safemath{\bmdelta}{\bideltad}

\safemath{\bA}{\mathbf{A}}
\safemath{\bB}{\mathbf{B}}
\safemath{\bC}{\mathbf{C}}
\safemath{\bD}{\mathbf{D}}
\safemath{\bE}{\mathbf{E}}
\safemath{\bF}{\mathbf{F}}
\safemath{\bG}{\mathbf{G}}
\safemath{\bH}{\mathbf{H}}
\safemath{\bI}{\mathbf{I}}
\safemath{\bJ}{\mathbf{J}}
\safemath{\bK}{\mathbf{K}}
\safemath{\bL}{\mathbf{L}}
\safemath{\bM}{\mathbf{M}}
\safemath{\bN}{\mathbf{N}}
\safemath{\bO}{\mathbf{O}}
\safemath{\bP}{\mathbf{P}}
\safemath{\bQ}{\mathbf{Q}}
\safemath{\bR}{\mathbf{R}}
\safemath{\bS}{\mathbf{S}}
\safemath{\bT}{\mathbf{T}}
\safemath{\bU}{\mathbf{U}}
\safemath{\bV}{\mathbf{V}}
\safemath{\bW}{\mathbf{W}}
\safemath{\bX}{\mathbf{X}}
\safemath{\bY}{\mathbf{Y}}
\safemath{\bZ}{\mathbf{Z}}

\safemath{\bZero}{\mathbf{0}}
\safemath{\bOne}{\mathbf{1}}
\safemath{\bDelta}{\mathbf{\Delta}}
\safemath{\bLambda}{\mathbf{\UpLambda}}
\safemath{\bPhi}{\mathbf{\Upphi}}
\safemath{\bSigma}{\mathbf{\Upsigma}}
\safemath{\bOmega}{\mathbf{\Upomega}}
\safemath{\bTheta}{\mathbf{\Uptheta}}

\bmdefine{\biAd}{A}
\bmdefine{\biBd}{B}
\bmdefine{\biCd}{C}
\bmdefine{\biDd}{D}
\bmdefine{\biEd}{E}
\bmdefine{\biFd}{F}
\bmdefine{\biGd}{G}
\bmdefine{\biHd}{H}
\bmdefine{\biId}{I}
\bmdefine{\biJd}{J}
\bmdefine{\biKd}{K}
\bmdefine{\biLd}{L}
\bmdefine{\biMd}{M}
\bmdefine{\biOd}{N}
\bmdefine{\biPd}{O}
\bmdefine{\biQd}{P}
\bmdefine{\biRd}{R}
\bmdefine{\biSd}{S}
\bmdefine{\biTd}{T}
\bmdefine{\biUd}{U}
\bmdefine{\biVd}{V}
\bmdefine{\biWd}{W}
\bmdefine{\biXd}{X}
\bmdefine{\biYd}{Y}
\bmdefine{\biZd}{Z}

\bmdefine{\biDelta}{\Delta}
\bmdefine{\biLambda}{\Lambda}
\bmdefine{\biPhi}{\Phi}
\bmdefine{\biSigma}{\Sigma}
\bmdefine{\biOmega}{\Omega}
\bmdefine{\biTheta}{\Theta}

\safemath{\bimA}{\biAd}
\safemath{\bimB}{\biBd}
\safemath{\bimC}{\biCd}
\safemath{\bimD}{\biDd}
\safemath{\bimE}{\biEd}
\safemath{\bimF}{\biFd}
\safemath{\bimG}{\biGd}
\safemath{\bimH}{\biHd}
\safemath{\bimI}{\biId}
\safemath{\bimJ}{\biJd}
\safemath{\bimK}{\biKd}
\safemath{\bimL}{\biLd}
\safemath{\bimM}{\biMd}
\safemath{\bimN}{\biNd}
\safemath{\bimO}{\biOd}
\safemath{\bimP}{\biPd}
\safemath{\bimQ}{\biQd}
\safemath{\bimR}{\biRd}
\safemath{\bimS}{\biSd}
\safemath{\bimT}{\biTd}
\safemath{\bimU}{\biUd}
\safemath{\bimV}{\biVd}
\safemath{\bimW}{\biWd}
\safemath{\bimX}{\biXd}
\safemath{\bimY}{\biYd}
\safemath{\bimZ}{\biZd}

\safemath{\bimDelta}{\biDelta}
\safemath{\bimLambda}{\biLambda}
\safemath{\bimPhi}{\biPhi}
\safemath{\bimSigma}{\biSigma}
\safemath{\bimOmega}{\biOmega}
\safemath{\bimTheta}{\biTheta}

\safemath{\setA}{\mathcal{A}}
\safemath{\setB}{\mathcal{B}}
\safemath{\setC}{\mathcal{C}}
\safemath{\setD}{\mathcal{D}}
\safemath{\setE}{\mathcal{E}}
\safemath{\setF}{\mathcal{F}}
\safemath{\setG}{\mathcal{G}}
\safemath{\setH}{\mathcal{H}}
\safemath{\setI}{\mathcal{I}}
\safemath{\setJ}{\mathcal{J}}
\safemath{\setK}{\mathcal{K}}
\safemath{\setL}{\mathcal{L}}
\safemath{\setM}{\mathcal{M}}
\safemath{\setN}{\mathcal{N}}
\safemath{\setO}{\mathcal{O}}
\safemath{\setP}{\mathcal{P}}
\safemath{\setQ}{\mathcal{Q}}
\safemath{\setR}{\mathcal{R}}
\safemath{\setS}{\mathcal{S}}
\safemath{\setT}{\mathcal{T}}
\safemath{\setU}{\mathcal{U}}
\safemath{\setV}{\mathcal{V}}
\safemath{\setW}{\mathcal{W}}
\safemath{\setX}{\mathcal{X}}
\safemath{\setY}{\mathcal{Y}}
\safemath{\setZ}{\mathcal{Z}}
\safemath{\emptySet}{\varnothing}

\safemath{\colA}{\mathscr{A}}
\safemath{\colB}{\mathscr{B}}
\safemath{\colC}{\mathscr{C}}
\safemath{\colD}{\mathscr{D}}
\safemath{\colE}{\mathscr{E}}
\safemath{\colF}{\mathscr{F}}
\safemath{\colG}{\mathscr{G}}
\safemath{\colH}{\mathscr{H}}
\safemath{\colI}{\mathscr{I}}
\safemath{\colJ}{\mathscr{J}}
\safemath{\colK}{\mathscr{K}}
\safemath{\colL}{\mathscr{L}}
\safemath{\colM}{\mathscr{M}}
\safemath{\colN}{\mathscr{N}}
\safemath{\colO}{\mathscr{O}}
\safemath{\colP}{\mathscr{P}}
\safemath{\colQ}{\mathscr{Q}}
\safemath{\colR}{\mathscr{R}}
\safemath{\colS}{\mathscr{S}}
\safemath{\colT}{\mathscr{T}}
\safemath{\colU}{\mathscr{U}}
\safemath{\colV}{\mathscr{V}}
\safemath{\colW}{\mathscr{W}}
\safemath{\colX}{\mathscr{X}}
\safemath{\colY}{\mathscr{Y}}
\safemath{\colZ}{\mathscr{Z}}

\safemath{\opA}{\mathbb{A}}
\safemath{\opB}{\mathbb{B}}
\safemath{\opC}{\mathbb{C}}
\safemath{\opD}{\mathbb{D}}
\safemath{\opE}{\mathbb{E}}
\safemath{\opF}{\mathbb{F}}
\safemath{\opG}{\mathbb{G}}
\safemath{\opH}{\mathbb{H}}
\safemath{\opI}{\mathbb{I}}
\safemath{\opJ}{\mathbb{J}}
\safemath{\opK}{\mathbb{K}}
\safemath{\opL}{\mathbb{L}}
\safemath{\opM}{\mathbb{M}}
\safemath{\opN}{\mathbb{N}}
\safemath{\opO}{\mathbb{O}}
\safemath{\opP}{\mathbb{P}}
\safemath{\opQ}{\mathbb{Q}}
\safemath{\opR}{\mathbb{R}}
\safemath{\opS}{\mathbb{S}}
\safemath{\opT}{\mathbb{T}}
\safemath{\opU}{\mathbb{U}}
\safemath{\opV}{\mathbb{V}}
\safemath{\opW}{\mathbb{W}}
\safemath{\opX}{\mathbb{X}}
\safemath{\opY}{\mathbb{Y}}
\safemath{\opZ}{\mathbb{Z}}
\safemath{\opZero}{\mathbb{O}}
\safemath{\identityop}{\opI}

\safemath{\veca}{\bma}
\safemath{\vecb}{\bmb}
\safemath{\vecc}{\bmc}
\safemath{\vecd}{\bmd}
\safemath{\vece}{\bme}
\safemath{\vecf}{\bmf}
\safemath{\vecg}{\bmg}
\safemath{\vech}{\bmh}
\safemath{\veci}{\bmi}
\safemath{\vecj}{\bmj}
\safemath{\veck}{\bmk}
\safemath{\vecl}{\bml}
\safemath{\vecm}{\bmm}
\safemath{\vecn}{\bmn}
\safemath{\veco}{\bmo}
\safemath{\vecp}{\bmp}
\safemath{\vecq}{\bmq}
\safemath{\vecr}{\bmr}
\safemath{\vecs}{\bms}
\safemath{\vect}{\bmt}
\safemath{\vecu}{\bmu}
\safemath{\vecv}{\bmv}
\safemath{\vecw}{\bmw}
\safemath{\vecx}{\bmx}
\safemath{\vecy}{\bmy}
\safemath{\vecz}{\bmz}

\safemath{\veczero}{\bmzero}
\safemath{\vecone}{\bmone}
\safemath{\vecxi}{\bmxi}
\safemath{\veclambda}{\bmlambda}
\safemath{\vecmu}{\bmmu}
\safemath{\vectheta}{\bmtheta}
\safemath{\vecphi}{\bmphi}
\safemath{\vecdelta}{\bmdelta}

\safemath{\matA}{\bA}
\safemath{\matB}{\bB}
\safemath{\matC}{\bC}
\safemath{\matD}{\bD}
\safemath{\matE}{\bE}
\safemath{\matF}{\bF}
\safemath{\matG}{\bG}
\safemath{\matH}{\bH}
\safemath{\matI}{\bI}
\safemath{\matJ}{\bJ}
\safemath{\matK}{\bK}
\safemath{\matL}{\bL}
\safemath{\matM}{\bM}
\safemath{\matN}{\bN}
\safemath{\matO}{\bO}
\safemath{\matP}{\bP}
\safemath{\matQ}{\bQ}
\safemath{\matR}{\bR}
\safemath{\matS}{\bS}
\safemath{\matT}{\bT}
\safemath{\matU}{\bU}
\safemath{\matV}{\bV}
\safemath{\matW}{\bW}
\safemath{\matX}{\bX}
\safemath{\matY}{\bY}
\safemath{\matZ}{\bZ}
\safemath{\matzero}{\bmzero}

\safemath{\matDelta}{\bDelta}
\safemath{\matLambda}{\bLambda}
\safemath{\matPhi}{\bPhi}
\safemath{\matSigma}{\bSigma}
\safemath{\matOmega}{\bOmega}
\safemath{\matTheta}{\bTheta}

\safemath{\matidentity}{\matI}
\safemath{\matone}{\matO}

\safemath{\rnda}{A}
\safemath{\rndb}{B}
\safemath{\rndc}{C}
\safemath{\rndd}{D}
\safemath{\rnde}{E}
\safemath{\rndf}{F}
\safemath{\rndg}{G}
\safemath{\rndh}{H}
\safemath{\rndi}{I}
\safemath{\rndj}{J}
\safemath{\rndk}{K}
\safemath{\rndl}{L}
\safemath{\rndm}{M}
\safemath{\rndn}{N}
\safemath{\rndo}{O}
\safemath{\rndp}{P}
\safemath{\rndq}{Q}
\safemath{\rndr}{R}
\safemath{\rnds}{S}
\safemath{\rndt}{T}
\safemath{\rndu}{U}
\safemath{\rndv}{V}
\safemath{\rndw}{W}
\safemath{\rndx}{X}
\safemath{\rndy}{Y}
\safemath{\rndz}{Z}

\safemath{\rveca}{\bimA}
\safemath{\rvecb}{\bimB}
\safemath{\rvecc}{\bimC}
\safemath{\rvecd}{\bimD}
\safemath{\rvece}{\bimE}
\safemath{\rvecf}{\bimF}
\safemath{\rvecg}{\bimG}
\safemath{\rvech}{\bimH}
\safemath{\rveci}{\bimI}
\safemath{\rvecj}{\bimJ}
\safemath{\rveck}{\bimK}
\safemath{\rvecl}{\bimL}
\safemath{\rvecm}{\bimM}
\safemath{\rvecn}{\bimN}
\safemath{\rveco}{\bomO}
\safemath{\rvecp}{\bimP}
\safemath{\rvecq}{\bimQ}
\safemath{\rvecr}{\bimR}
\safemath{\rvecs}{\bimS}
\safemath{\rvect}{\bimT}
\safemath{\rvecu}{\bimU}
\safemath{\rvecv}{\bimV}
\safemath{\rvecw}{\bimW}
\safemath{\rvecx}{\bimX}
\safemath{\rvecy}{\bimY}
\safemath{\rvecz}{\bimZ}

\safemath{\rvecxi}{\bmxi}
\safemath{\rveclambda}{\bmlambda}
\safemath{\rvecmu}{\bmmu}
\safemath{\rvectheta}{\bmtheta}
\safemath{\rvecphi}{\bmphi}

\safemath{\rmatA}{\bimA}
\safemath{\rmatB}{\bimB}
\safemath{\rmatC}{\bimC}
\safemath{\rmatD}{\bimD}
\safemath{\rmatE}{\bimE}
\safemath{\rmatF}{\bimF}
\safemath{\rmatG}{\bimG}
\safemath{\rmatH}{\bimH}
\safemath{\rmatI}{\bimI}
\safemath{\rmatJ}{\bimJ}
\safemath{\rmatK}{\bimK}
\safemath{\rmatL}{\bimL}
\safemath{\rmatM}{\bimM}
\safemath{\rmatN}{\bimN}
\safemath{\rmatO}{\bimO}
\safemath{\rmatP}{\bimP}
\safemath{\rmatQ}{\bimQ}
\safemath{\rmatR}{\bimR}
\safemath{\rmatS}{\bimS}
\safemath{\rmatT}{\bimT}
\safemath{\rmatU}{\bimU}
\safemath{\rmatV}{\bimV}
\safemath{\rmatW}{\bimW}
\safemath{\rmatX}{\bimX}
\safemath{\rmatY}{\bimY}
\safemath{\rmatZ}{\bimZ}

\safemath{\rmatDelta}{\bimDelta}
\safemath{\rmatLambda}{\bimLambda}
\safemath{\rmatPhi}{\bimPhi}
\safemath{\rmatSigma}{\bimSigma}
\safemath{\rmatOmega}{\bimOmega}
\safemath{\rmatTheta}{\bimTheta}

\usepackage{amssymb}
\usepackage{amsfonts}
\usepackage{mathrsfs}
\usepackage{xspace}
\usepackage{bm}
\usepackage{fancyref}
\usepackage{textcomp}

\usepackage{multirow}
\usepackage{stmaryrd}

\newenvironment{textbmatrix}{	\setlength{\arraycolsep}{2.5pt}%
								\big[\begin{matrix}}{\end{matrix}\big]%
								\raisebox{0.08ex}{\vphantom{M}}}

\def\be{\begin{equation}}
\def\ee{\end{equation}}
\def\een{\nonumber \end{equation}}
\def\mat{\begin{bmatrix}}
\def\emat{\end{bmatrix}}
\def\btm{\begin{textbmatrix}}
\def\etm{\end{textbmatrix}}

\def\ba#1\ea{\begin{align}#1\end{align}}
\def\bas#1\eas{\begin{align*}#1\end{align*}}
\def\bs#1\es{\begin{split}#1\end{split}} 
\def\bg#1\eg{\begin{gather}#1\end{gather}}
\def\bml#1\eml{\begin{multline}#1\end{multline}}
\def\bi#1\ei{\begin{itemize}#1\end{itemize}}

\newcommand{\lefto}{\mathopen{}\left}

\newcommand{\card}[1]{\lvert#1\rvert}			

\newcommand{\vecnorm}[1]{\lefto\lVert#1\right\rVert}		
\newcommand{\opnorm}[1]{\lVert#1\rVert}		

\safemath{\dirac}{\delta}					
\safemath{\krond}{\dirac}					

\safemath{\upto}{\uparrow}
\safemath{\downto}{\downarrow}
\safemath{\iu}{j}							
\safemath{\ev}{\lambda}						
\safemath{\hilseqspace}{l^{2}}				
\newcommand{\banachfunspace}[1]{\setL^{#1}}	
\safemath{\hilfunspace}{\banachfunspace{2}}	

\safemath{\SNR}{\textsf{SNR}} 				
\safemath{\PAR}{\textsf{PAR}} 				
\safemath{\No}{N_0}							
\safemath{\Es}{E_s}							
\safemath{\Eb}{E_b}							
\safemath{\EbNo}{\frac{\Eb}{\No}}
\safemath{\EsNo}{\frac{\Es}{\No}}

\DeclareMathOperator{\CHop}{\ensuremath{\opH}} 
\safemath{\tvir}{\rndh_{\CHop}}				
\safemath{\tvtf}{\rndl_{\CHop}}				
\safemath{\spf}{\rnds_{\CHop}}				
\safemath{\bff}{H_{\CHop}}					

\safemath{\ircf}{r_{h}}						
\safemath{\tftvcf}{r_{s}}					
\safemath{\tfcf}{r_{l}}						
\safemath{\bfcf}{r_{H}}						

\safemath{\tcorr}{c_h}						
\safemath{\scf}{c_{s}}						
\safemath{\tfcorr}{c_{l}}					
\safemath{\fcorr}{c_{H}}						

\safemath{\mi}{I}							
\safemath{\capacity}{C}						

\safemath{\normal}{\mathcal{N}}			
\safemath{\jpg}{\mathcal{CN}}			
\safemath{\mchain}{\leftrightarrow}		

\safemath{\dB}{\,\mathrm{dB}}
\safemath{\dBm}{\,\mathrm{dBm}}
\safemath{\Hz}{\,\mathrm{Hz}}
\safemath{\kHz}{\,\mathrm{kHz}}
\safemath{\MHz}{\,\mathrm{MHz}}
\safemath{\GHz}{\,\mathrm{GHz}}
\safemath{\s}{\,\mathrm{s}}
\safemath{\ms}{\,\mathrm{ms}}
\safemath{\mus}{\,\mathrm{\text{\textmu}s}}
\safemath{\ns}{\,\mathrm{ns}}
\safemath{\ps}{\,\mathrm{ps}}
\safemath{\meter}{\,\mathrm{m}}
\safemath{\mm}{\,\mathrm{mm}}
\safemath{\cm}{\,\mathrm{cm}}
\safemath{\m}{\,\mathrm{m}}
\safemath{\W}{\,\mathrm{W}}
\safemath{\mW}{\, \mathrm{mW}}
\safemath{\J}{\,\mathrm{J}}
\safemath{\K}{\,\mathrm{K}}
\safemath{\bit}{\,\mathrm{bit}}
\safemath{\nat}{\,\mathrm{nat}}

\safemath{\define}{\triangleq}			

\safemath{\equivalent}{\sim}
\safemath{\distas}{\sim}					
\safemath{\sdiff}{\Delta}				

\safemath{\reals}{\mathbb{R}}
\safemath{\positivereals}{\reals_{+}}
\safemath{\integers}{\mathbb{Z}}
\safemath{\posint}{\integers_{+}}
\safemath{\naturals}{\mathbb{N}}
\safemath{\posnaturals}{\naturals_{+}}
\safemath{\complexset}{\mathbb{C}}
\safemath{\rationals}{\mathbb{Q}}

\newcommand*{\fancyrefapplabelprefix}{app}		
\newcommand*{\fancyrefthmlabelprefix}{thm}		
\newcommand*{\fancyreflemlabelprefix}{lem}		
\newcommand*{\fancyrefcorlabelprefix}{cor}		
\newcommand*{\fancyrefdeflabelprefix}{def}		
\newcommand*{\fancyrefproplabelprefix}{prop}	
\newcommand*{\fancyrefobslabelprefix}{obs}		
\newcommand*{\fancyrefalglabelprefix}{alg}		
\newcommand*{\fancyrefremlabelprefix}{rem}		
\newcommand*{\fancyrefasmlabelprefix}{asm}	    

\frefformat{vario}{\fancyrefseclabelprefix}{Section~#1}
\frefformat{vario}{\fancyrefthmlabelprefix}{Theorem~#1}
\frefformat{vario}{\fancyreflemlabelprefix}{Lemma~#1}
\frefformat{vario}{\fancyrefcorlabelprefix}{Corollary~#1}
\frefformat{vario}{\fancyrefdeflabelprefix}{Definition~#1}
\frefformat{vario}{\fancyrefobslabelprefix}{Observation~#1}
\frefformat{vario}{\fancyrefasmlabelprefix}{Assumption~#1}
\frefformat{vario}{\fancyreffiglabelprefix}{Fig.~#1}
\frefformat{vario}{\fancyrefapplabelprefix}{Appendix~#1} 
\frefformat{vario}{\fancyrefproplabelprefix}{Proposition~#1}
\frefformat{vario}{\fancyrefalglabelprefix}{Algorithm~#1}
\frefformat{vario}{\fancyrefremlabelprefix}{Remark~#1}
\frefformat{vario}{\fancyrefeqlabelprefix}{(#1)}

\newtheorem{rem}{Remark}

\safemath{\dictab}{[\,\dicta\,\,\dictb\,]}

\safemath{\ysig}{\bmy}
\safemath{\ysighat}{\hat{\ysig}}
\safemath{\ysigdim}{M}
\safemath{\xsig}{\bmx}
\safemath{\xsigdim}{N}
\safemath{\nx}{n_x}
\safemath{\zsig}{\bmz}
\safemath{\zsigdim}{\ysigdim}
\safemath{\rsig}{\bmr}
\safemath{\Adict}{\bA}
\safemath{\Adicttilde}{\widetilde{\Adict}}
\safemath{\Adictdim}{\outputdim\times\xsigdim}
\safemath{\avec}{\bma}
\safemath{\avectilde}{\tilde{\avec}}
\safemath{\Bdict}{\bB}
\safemath{\Bdicttilde}{\widetilde{\Bdict}}
\safemath{\Cdict}{\bC}
\safemath{\cvec}{\bmc}
\safemath{\Ddict}{\bD}
\safemath{\Ddictdim}{\ysigdim\times\xsigdim}
\safemath{\dvec}{\bmd}
\safemath{\Ddicttilde}{\widetilde{\bD}}
\safemath{\Bonb}{\bB}
\safemath{\bvec}{\bmb}
\safemath{\Bonbdim}{\ysigdim\times\ysigdim}
\safemath{\noise}{\bmn}
\safemath{\noisedim}{\ysigim}
\safemath{\err}{\bme}
\safemath{\errdim}{\ysigdim}
\safemath{\errset}{\setE}
\safemath{\nerr}{n_e}
\safemath{\delop}{\bP_\errset}
\safemath{\delopc}{\bP_{{\errset}^c}}

\safemath{\cplxi}{\imath}
\safemath{\cplxj}{\jmath}

\safemath{\dict}{\matD}
\safemath{\inputdim}{N}		
\safemath{\outputdim}{M}		
\safemath{\sparsity}{S}	
\safemath{\inputdimA}{{N_a}}	
\safemath{\inputdimB}{{N_b}}	
\safemath{\elemA}{{n_a}}	
\safemath{\elemB}{{n_b}}	
\safemath{\resA}{\matR_a}	
\safemath{\resB}{\matR_b}	
\safemath{\subD}{\matS} 
\safemath{\subA}{\matS_a} 
\safemath{\subB}{\matS_b} 
\safemath{\dicta}{\matA} 	
\safemath{\dictb}{\matB} 	
\safemath{\hollowS}{H}
\safemath{\hollowA}{H_a}
\safemath{\hollowB}{H_b}
\safemath{\cross}{Z}
\safemath{\coh}{\mu_d}			
\safemath{\coha}{\mu_a}			
\safemath{\cohb}{\mu_b}			
\safemath{\mubs}{\nu}	
\safemath{\cohm}{\mu_m} 
\safemath{\dictset}{\setD}	
\safemath{\dictsetp}{\dictset(\coh,\coha,\cohb)}	
\safemath{\dictsetgen}{\dictset_\text{gen}}
\safemath{\dictsetgenp}{\dictsetgen(\coh)}
\safemath{\dictsetonb}{\dictset_\text{onb}}
\safemath{\dictsetonbp}{\dictsetonb(\coh)}

\safemath{\leftside}{U}
\safemath{\rightsideA}{R_a}
\safemath{\rightsideB}{R_b}

\safemath{\indexS}{\setI_S} 

\safemath{\na}{n_a}			
\safemath{\nb}{n_b}			
\safemath{\coeffa}{p_i}	
\safemath{\coeffb}{q_j}	
\safemath{\seta}{\setP}		
\safemath{\setb}{\setQ}     
\safemath{\setw}{\setW}	
\safemath{\setz}{\setZ}	
\safemath{\cola}{\veca}		
\safemath{\colb}{\vecb}		
\safemath{\cold}{\vecd}		
\safemath{\inputvec}{\vecx} 	
\safemath{\error}{\vece}	
\safemath{\noiseout}{\vecz} 	
\safemath{\inputvecel}{x}
\safemath{\inputveca}{\vecx_a}
\safemath{\inputvecb}{\vecx_b}
\safemath{\outputvec}{\vecy}	
\safemath{\lambdamin}{\lambda_{\mathrm{min}}}

\safemath{\elltwo}{\ell_2}
\safemath{\ellone}{\ell_1}
\safemath{\ellzero}{\ell_0}
\safemath{\ellinf}{\ell_\infty}
\safemath{\ellinftilde}{\ell_{\widetilde\infty}}
\safemath{\licard}{Z(\coh,\coha,\cohb)}
\safemath{\xsol}{\hat{x}}
\safemath{\xbord}{x_b}		
\safemath{\xstat}{x_s}		
\safemath{\xstatLone}{\tilde{x}_s}
\safemath{\order}{\mathcal{O}} 
\safemath{\scales}{\Theta} 
\safemath{\ones}{\mathbf{1}} 
\safemath{\zeroes}{\mathbf{0}} 
\safemath{\thlone}{\kappa(\coh,\cohb)} 
\safemath{\constoneA}{\delta} 
\safemath{\constoneB}{\epsilon} 
\safemath{\nlarge}{L}				   
\safemath{\sumlarge}{S_\nlarge}
\safemath{\maxlarger}{P_\nlarge}	   
\safemath{\Pzero}{\textrm{P0}}	
\safemath{\Pone}{\textrm{P1}}
\safemath{\vecfir}{\vecw}			 
\safemath{\vecsec}{\vecz}
\safemath{\elvecfir}{w}              
\safemath{\elvecsec}{z}				 
\safemath{\nlargefir}{n}
\safemath{\normout}{\gamma}
\safemath{\auxfun}{h}
\safemath{\supp}{\textrm{supp}}

\safemath{\indexa}{\ell}
\safemath{\indexb}{r}
\safemath{\indexc}{i}
\safemath{\indexd}{j}

\safemath{\project}{P}

\allowdisplaybreaks

\begin{document}

%
\title{MSE-Optimal 1-Bit Precoding for Multiuser MIMO via Branch and Bound }

\name{Sven Jacobsson$^{\pumpkin,\mathghost,\mathbat}$, Weiyu Xu$^{\mathwitch}$, Giuseppe Durisi$^{\mathghost}$, and Christoph Studer$^{\pumpkin}$\thanks{The work of SJ and GD was supported in part by the Swedish Foundation for Strategic Research under grant ID14-0022, and by the Swedish Governmental Agency for Innovation Systems (VINNOVA) within the competence center ChaseOn. SJ's research visit at Cornell was sponsored in part by Cornell's College of Engineering. The work of CS was supported in part by Xilinx, Inc.~and by the US National Science Foundation~(NSF) under grants ECCS-1408006, CCF-1535897, CAREER CCF-1652065, and CNS-1717559.}
}
\address{
$^{\pumpkin}$Cornell University, Ithaca, NY, USA; 
$^{\mathghost}$Chalmers University of Technology, Gothenburg, Sweden;  \\
$^{\mathbat}$Ericsson Research, Gothenburg, Sweden;
$^{\mathwitch}$The University of Iowa, Iowa City, IA, USA}

\maketitle


\vspace{-1cm}
\begin{abstract}

In this paper, we solve the sum mean-squared error (MSE)-optimal 1-bit quantized precoding problem exactly for small-to-moderate sized multiuser multiple-input multiple-output (MU-MIMO) systems via branch and bound.
To this end, we reformulate the original NP-hard precoding problem as a tree search and deploy a number of strategies that improve the pruning efficiency without sacrificing optimality. 
We evaluate the error-rate performance and the complexity of the resulting 1-bit branch-and-bound (BB-1) precoder, and compare its efficacy to that of existing, suboptimal algorithms for 1-bit precoding in MU-MIMO systems.
\end{abstract}



\section{Introduction}

Massive multiuser multiple-input multiple-output (MU-MIMO) technology, a scaled-up version of what is used in today's cellular communication systems,  is expected to play a critical role in next-generation wireless systems~\cite{boccardi14a}.
In the downlink, the basestation (BS) transmits data to multiple users in the same time-frequency resource by mapping the information symbols to the antenna array via a precoder~\cite{spencer04a,rusek14a}.
For massive MU-MIMO systems, the increase in the number of BS antenna elements entails significant growths in circuit power consumption and interconnect bandwidth over the link connecting the baseband processing unit to the radio unit. These challenges are further aggravated when operating over large bandwidths at millimeter-wave frequencies~\cite{heath-jr.15a}. 

The use of low-resolution digital-to-analog converters (DACs) has recently been proposed to reduce power consumption and mitigate the interconnect-bandwidth bottleneck at the BS. 
When low-resolution DACs are used, each entry of the precoded vector must be quantized to the low-cardinality alphabet that is supported by the transcoder in the DAC. For the special case of 1-bit DACs and frequency-flat channels, 1-bit precoding has been studied in, e.g., \cite{mezghani09c, jedda16a, swindlehurst17a, landau17a, jacobsson17d, castaneda17a, li17a, tirkkonen17a}; the frequency-selective scenario has been studied recently in~\cite{jacobsson17f, jacobsson17c}. 

Unfortunately, the mean-squared error (MSE)-optimal precoding problem for the case where the precoded vector is quantized to a finite alphabet is, in general, NP-hard. 
By relaxing the finite-alphabet constraint to a convex set, suboptimal precoders (with near-optimal performance) have been developed~\cite{jacobsson17d, castaneda17a,jacobsson17f}. 
%
%
In this paper, we solve the sum MSE-optimal 1-bit precoding \emph{exactly} for small-to-moderate sized MU-MIMO systems (e.g., $12$ BS antennas) that operate over frequency-flat channels without resorting to an exhaustive search. To this end, we reformulate the NP-hard  problem as a tree search, which we then solve by the proposed 1-bit branch-and-bound (\mbox{BB-1})~precoding algorithm. We deploy a number of strategies that improve the pruning efficiency of BB-1 without sacrificing optimality, and we compare the error-rate performance and complexity of BB-1 to that of existing, suboptimal~precoders.

\section{MSE-Optimal Quantized Precoding}

\subsection{The Quantized Precoding (QP) Problem}

We consider quantized (or finite-alphabet) precoding for the MU-MIMO downlink. The BS is equipped with $B$ antennas and serves $U$ single-antenna users in the same time-frequency resource.
%
%
The goal of the sum MSE-optimal quantized precoder is to compute a precoded vector $\bmx\in\setX^B$, with $\setX$ being the finite-cardinality transmit alphabet, by solving the following quantized precoding (QP) problem \cite{jacobsson17d}:
\begin{align*} 
\text{(QP)}\quad \left\{\begin{array}{ll}
\underset{\bmx \in \setX^{B},\, \beta \in \reals}{\textrm{minimize}} & \vecnorm{\vecs -  \beta\matH\vecx}^2_2 + \beta^2 U N_0 \\
\,\textrm{subject to} & \|\bmx\|_2^2\leq 1 \text{ and } \beta > 0.
\end{array}\right.
\end{align*}
Here, $\bms\in\setO^U$ is the (known) data vector to be transmitted to the $U$ users, $\setO$ is the constellation alphabet (e.g., 16-QAM), $\bH\in\complexset^{U\times B}$ is the (known) downlink channel matrix, and $\No$ is the noise variance at each user (assumed to be equal for all users and known at the BS). 
We define the signal-to-noise ratio (SNR) as $\textit{SNR} = 1/N_0$.
The \emph{precoding factor} $\beta$ takes into account the gain of the channel~\cite{jacobsson17d}.
We note that for a given value of~$\beta$, the (QP) problem is a closest vector problem~(CVP) that is NP hard~\cite{verdu89a}. As a consequence, solving (QP) via an exhaustive search requires evaluating the objective function in~(QP) for $\card{\setX}^B$ candidate vectors, which is infeasible for moderate-to-large~$B$.
Hence, more efficient precoding algorithms are~required in practice.

\subsection{Rewriting the (QP) Problem}
We start by using the fact that the precoding factor $\beta>0$ in~(QP) is a continuous parameter. 
Hence, given a precoded vector $\bmx$ for which $\Re\{\bmx^H\bH^H\bms\}>0$, the optimal associated precoding factor can be readily computed as
\begin{align*} 
\hat\beta(\vecx)
= \frac{ \Re\{\bmx^H\bH^H\bms\}}{\|\bH\bmx\|_2^2 + \No U}.
\end{align*}
By inserting this optimal precoding factor $\hat\beta(\bmx)$ into the objective function of (QP), we obtain
\begin{align*}
 \opnorm{\vecs - \hat\beta(\vecx) \matH\vecx}^2_2 + \hat\beta(\vecx)^2 U N_0= 
 \opnorm{\vecs}^2 - \frac{ \Re\{\bmx^H\bH^H\bms\}^2}{\opnorm{\bH\bmx}_2^2 + \No U}.
\end{align*}
Consequently, solving the problem (QP) is equivalent to solving the following optimization problem:
\begin{align*} 
\text{(QP$^*$)}\quad \left\{\begin{array}{ll}
\underset{\bmx \in \setX^{B}}{\textrm{minimize}} & \displaystyle \frac{\|\bH\bmx\|_2^2 + \No U}{ \Re\{\bmx^H\bH^H\bms\}^2} \\[0.3cm]
\textrm{subject to} & \|\bmx\|_2^2\leq 1.
\end{array}\right.
\end{align*}
Let $\hat\vecx$ denote the optimal solution to the problem (QP$^*$). Note that the corresponding precoding factor $\hat{\beta}(\hat\vecx)$ can be negative. In this case, we use that $\hat{\beta}(-\hat\vecx) = -\hat{\beta}(\hat\vecx)$ to simply flip the sign of the solution $\hat\bmx$.
For this not to affect the solution, we require that $\setX$ is {symmetric}, i.e., that $x\in \setX$ implies $-x \in \setX$. 

\section{BB-1: 1-bit Branch-and-Bound Precoder}
\subsection{Simplifying (QP*) for Constant-Modulus Alphabets}

To arrive at a formulation of (QP*) that is amenable to branch and bound, we triangularize the problem. For this to work, we require constant-modulus (CM) transmit alphabets $\setX$, i.e., that $|x|^2=1/B$ for all $x\in\setX$, which implies that $\|\bmx\|^2_2=1$. We use this property to rewrite the objective of (QP*) as follows:
\begin{align*} 
\frac{\|\bH\bmx\|_2^2 + \No U}{\Re\{\bmx^H\bH^H\bms\}^2} = \frac{\|\bH\bmx\|_2^2 + {\No U}\|\bmx\|_2^2}{ \Re\{\bmx^H\bH^H\bms\}^2} = \frac{\|\widetilde\bH\bmx\|_2^2}{ \Re\{\bmx^H\bmz^\text{MRT}\}^2}.
\end{align*}
Here, $\widetilde\bH=\big[\bH^T, \sqrt{\No U}\,\bI_B \big]^T$ is the $(U+B)\times B$ augmented channel matrix and $\vecz^\text{MRT}= \bH^H \vecs$ is the $B$-dimensional maximal-ratio transmission (MRT) vector. By applying the QR factorization $\widetilde\bH=\bQ\bR$, where~$\bQ \in \opC^{(U+B)\times B}$ has unitary rows and $\bR\in\complexset^{B\times B}$ is an upper-triangular matrix with nonnegative values on the main diagonal, we can formulate the CM quantized precoding problem~as
\begin{align*} 
\text{(CMQP)}\quad  
\underset{\bmx \in \setX^{B}}{\textrm{minimize}} \,\,  \displaystyle \frac{\|\bR\bmx\|_2^2 }{ \Re\{\bmx^H \bmz^\text{MRT}\}^2}, 
\end{align*}
which is sum MSE-optimal for CM transmit alphabets~$\setX$.

\begin{figure}
\centering
\includegraphics[width = .9\columnwidth]{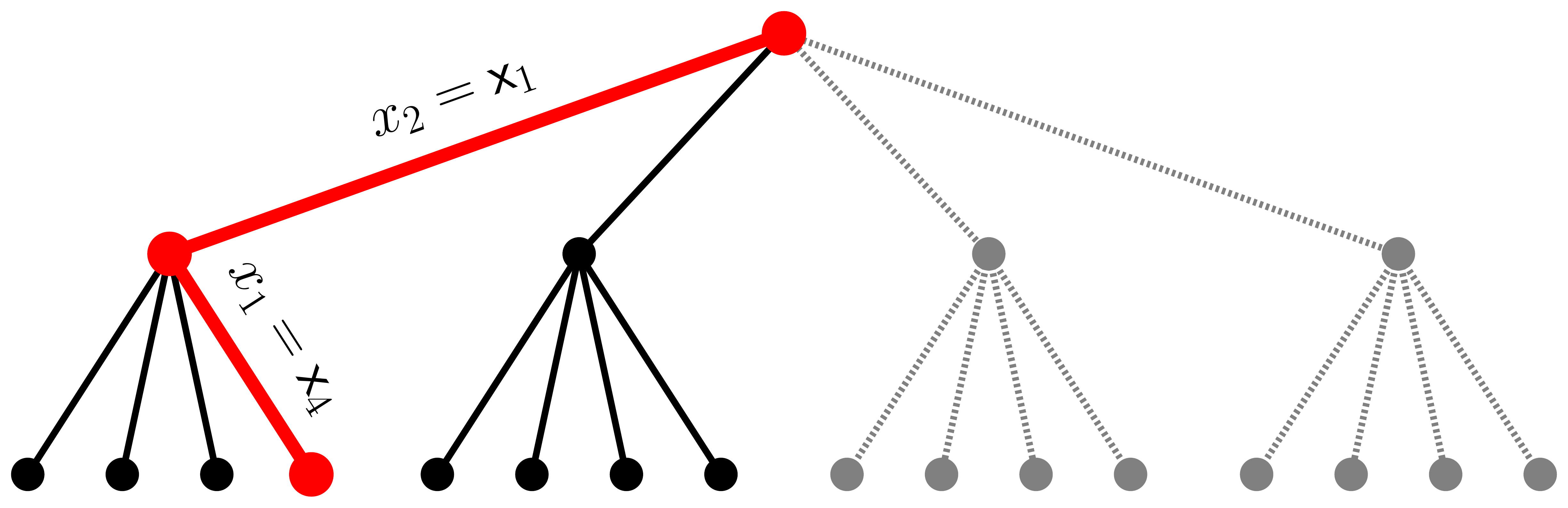}
\caption{Quaternary tree for the 1-bit-quantized case and for $B=2$. Note that the root of the tree corresponds to the last entry of the vector $\vecx$, and that half of the tree can by prepruned by symmetry (see~\fref{sec:preprune}). The path highlighted in red corresponds to the precoded vector $\vecx = [\mathsf{x}_4, \mathsf{x}_1 ]^T$. }	
\label{fig:tree}
\end{figure}

\subsection{Branch-and-Bound Procedure} \label{sec:BnB}

The branch-and-bound procedure proposed in this paper finds the optimal solution $\hat\bmx \in \setX^B$ to the problem (CMQP).
For the sake of brevity, we will focus exclusively on the 1-bit quantized case, where $\setX = \{\mathsf{x}_1, \mathsf{x}_2, \mathsf{x}_3, \mathsf{x}_4 \}$ and $\mathsf{x}_m = \frac{1}{\sqrt{B}}e^{j\pi( \frac{m}{2} - \frac{1}{4})}$.

The goal of using branch and bound to solve the problem (CMQP) is to reformulate it as a tree-search problem for which we can prune large parts of the tree in order to reduce the computational complexity.
It is key to realize that the problem (CMQP) can be associated with a $B$-level $|\setX|$-tree; see \fref{fig:tree} for an illustration. 
Each node in the tree at level $L$ can be uniquely described by the partial symbol vector~(PSV) $\bmx^{(L)} = [x_{L}, x_{L+1}, \dots, x_{B}]^T \in \setX^{B-L+1}$.
Consider branching out from a node in the tree at level $L+1$. The goals are to decide (i) which child node should be visited next and (ii) which child nodes can be pruned. 
To this end, we need a cost that represents the objective function in (CMQP) given a previously chosen PSV $\bmx^{(L+1)}$ and the potential child nodes $\tilde{x}\in\setX$ so that we can prune whenever the cost associated with a node at level $L$ exceeds some bound.
With this in mind, we lower-bound the numerator of the objective function in (CMQP) as
\begin{align*}
\|\bR\bmx\|^2 = \textstyle\sum_{b=1}^{B}\lefto|\textstyle\sum_{k=b}^B R_{b,k} x_k\right|^2 \geq n_L\big(\tilde{x};\bmx^{(L+1)}\big),
\end{align*}
where $n_L(\tilde{x};\bmx^{(L+1)})\geq0$ depends only on the potential child node $\tilde{x}\in\setX$ and the PSV $\bmx^{(L+1)}$. 
Similarly, we upper-bound the denominator of the objective function in (CMQP)~as
\begin{align*}
\Re\{\bmx^H \bmz^\text{MRT}\}^2 = \left(\textstyle\sum_{b=1}^{B} \Re\lefto\{ x_b^* z^\text{MRT}_b  \right\}\right)^2 \leq d_L\big(\tilde{x};\bmx^{(L+1)}\big),
\end{align*}
where $z^\text{MRT}_b$ is the $b$th entry of $\bmz^\text{MRT}$ and $d_L(\tilde{x};\bmx^{(L+1)})\geq0$ depends only on the potential child node $\tilde{x}\in\setX$ and the PSV $\bmx^{(L+1)}$. 
Given these two quantities we can define the cost $c_L(\tilde{x};\bmx^{(L+1)})$~as follows:
\begin{align*}
c_L\big(\tilde{x};\bmx^{(L+1)}\big)  = \frac{n_L\big(\tilde{x};\bmx^{(L+1)}\big) }{d_L\big(\tilde{x};\bmx^{(L+1)}\big)} \leq \frac{\|\bR\bmx\|^2}{\Re\{\bmx^H\bmz^\text{MRT}\}^2}.
\end{align*}
Whenever $c_L(\tilde{x};\bmx^{(L+1)})> \rho$, where $\rho>0$ is a suitably chosen treshold (we shall discuss how to set $\rho$ in Sections~\ref{sec:traversal} and \ref{sec:radius_init}), we can prune the corresponding subtree. Next, we provide specific choices for $n_L(\tilde{x};\bmx^{(L+1)})$ and $d_L(\tilde{x};\bmx^{(L+1)})$.

%
\subsection{Bounding The Cost Function}
Given a PSV $\bmx^{(L+1)}\in\setX^{B-L}$ and a candidate child $\tilde{x}$, we write the numerator of the cost function $c_L(\cdot,\cdot)$ as a sum of three parts---past, present, and future---as follows:
\begin{align*}
n_L\big(\tilde{x};\bmx^{(L+1)}\big) &= n_L^\text{past}(\bmx^{(L+1)}) + n_L^\text{present}\big(\tilde{x};\bmx^{(L+1)}\big) \\
& \quad + n_L^\text{future}\big(\tilde{x};\bmx^{(L+1)}\big).
\end{align*}
The past  is determined by the previously chosen PSV $\bmx^{(L+1)}$ and is given~by $n_L^\text{past}(\bmx^{(L+1)}) = \sum_{b=L+1}^{B}\big|\sum_{\ell=b}^B R_{b,\ell} x_\ell\big|^2$.
The present  depends on the choice of the child node $\tilde{x} \in \setX$ and on the PSV $\bmx^{(L+1)}$ and is given by $n_L^\text{present}(\tilde{x};\bmx^{(L+1)}) = \big|R_{L,L}\tilde{x} + \sum_{\ell=L+1}^B R_{L,\ell} x_\ell\big|^2$.
The future depends on the cost of all possible leaf nodes and is given by
\begin{align*}
n_L^\text{future}\big(\tilde{x};\bmx^{(L+1)}\big) &= \min_{\bar\bmx\in\setX^{L-1}} \textstyle\sum_{b=1}^{L-1} \Big|\! \textstyle\sum_{\ell=b}^{L-1} R_{b,\ell} \bar{x}_\ell \\
& \quad + R_{b,L}\tilde{x} + \textstyle\sum_{\ell=L+1}^B R_{b,\ell} x_\ell \Big|^2.
\end{align*}
Unfortunately, computing the future cost exactly is as hard as solving the original precoding problem. A trivial lower bound is obtained by setting $n_L^\text{future}(\tilde{x};\bmx^{(L+1)})=0$, which results, however, in a poor pruning behavior. In~\fref{sec:future}, we provide a more sophisticated approach that improves the pruning behavior.

Using a similar approach, we decompose the denominator of the cost associated with branching out from a node at level $L+1$ into three parts: past, present, and future. To arrive at an upper bound on the denominator of the cost function, we use the triangle inequality to bound $\big(\!\sum_{b=1}^{B} \Re\big\{ x_b^* z^\text{MRT}_b  \big\}\big)^2$ by
\begin{align*}
d_L(\tilde{x};\bmx^{(L+1)}) &=  \Big( |d_L^\text{\,past}(\bmx^{(L+1)}) + d_L^\text{\,present}(\tilde{x})\big| + \big|d_L^\text{\,future}\big|  \Big)^2\!,
\end{align*}
where the past and present are given by $d_L^\text{\,past}(\bmx^{(L+1)}) = \textstyle\sum_{\ell=L+1}^{B} \Re\{x_\ell^* z^\text{MRT}_\ell \}$ and $d_L^\text{\,present}(\tilde{x}) = \Re\{\tilde{x}^* z^\text{MRT}_L \}$, respectively. Finally, the future cost is
\begin{align*}
d_L^\text{\,future}  & = \max_{\bar{\bmx} \in \setX^{L-1}} \textstyle\sum_{\ell=1}^{L-1} \Re\big\{ \bar{x}_\ell^* z^\text{MRT}_\ell \big\}. 
\end{align*}
It can be shown that the maximum is achieved by $\bar{x}_\ell = x^\text{MRT}_\ell$, where $x^\text{MRT}_\ell = \text{argmin}_{x \in \setX} \lvert{x - z_\ell^\text{MRT}}\rvert^2$.

\section{Five Tricks that Make BB-1 Faster} \label{sec:tricks}
We now propose five tricks that improve the pruning behavior of the proposed algorithm without sacrificing optimality.

\subsection{Trick 1: Depth-First Best-First Tree Traversal with Radius Reduction} \label{sec:traversal}

We traverse the search tree in the following manner: at level $L+1$, we pick the $\tilde{x}$ that minimizes the current cost $c_L(\tilde{x}; \bmx^{(L+1)})$; we then proceed in a depth-first manner. Whenever a valid leaf node~$\bmx^{(1)}$ is found, we update the radius (bound) to
\begin{align*}
\rho \gets  \frac{\|\bR\bmx^{(1)}\|_2^2 }{ \Re\big\{(\bmx^{(1)})^H\bmz^\text{MRT}\big\}^2} 
\end{align*}
and we perform backtracking by proceeding upwards and selecting the next-best symbol, excluding branches that have been  explored or with a cost that exceeds the new radius. 

\begin{rem}
Any other tree-traversal strategy could be used, such as breadth-first used in the (suboptimal) K-best algorithm that can be implemented efficiently in hardware~\cite{wong02a, wenk06a}.
\end{rem}

\subsection{Trick 2: Radius Initialization} \label{sec:radius_init}

The pruning efficiency can be improved by initializing the tree search with some radius $\rho < \infty$, which is sufficiently large not to  exclude the optimal solution~\cite{burg05a}. 
We initialize the radius using the Wiener-filter (WF) solution, which can be computed at low complexity and can be shown to be optimal in the low-SNR regime.
Specifically, we initialize
\begin{align*}
	\rho &=  \frac{\opnorm{\bR\bmx^\text{WF}}_2^2}{\Re\big\{ \lefto( \vecx^\text{WF} \right)^H \vecz^\text{MRT}\big\}^2}.
\end{align*}
Here, the $b$th entry of $\bmx^\text{WF}$ is $x_b^\text{WF} = \text{argmin}_{x \in \setX} \lvert{x - z_b^\text{WF}}\rvert^2$ where $z_b^\text{WF}$ is the $b$th entry of $\vecz^\text{WF} = \matH^H(\matH\matH^H$ $+ U N_0\matI_U)^{-1}$.


\subsection{Trick 3: Sorted QR Decomposition} \label{sec:sorted}

%
%
We permute the columns of $\widetilde\matH$ (and the corresponding entries in $\vecx$) using the sorted-QR-decomposition algorithm put forward in~\cite{wubben01a}, so that the diagonal elements of $\bR$ are sorted in ascending order. This approach  improves substantially the pruning behavior for nodes close to the root because larger part of the search tree can be pruned early on.

\subsection{Trick 4: Predicting the Future}
\label{sec:future}

The pruning efficiency can further be improved by finding a nontrivial lower bound on $n_L^\text{future}(x_L;\bmx^{(L+1)})$. 
%
%
We denote by $\bR_{L-1} \in \opC^{L-1 \times L-1}$ the submatrix of $\bR$ whose entry on the $b$th row ($b = 1,2,\dots,L-1$) and on the $\ell$th column ($\ell = 1,2,\dots,L-1$) is~$R_{b,\ell}$
Furthermore, we denote by $\bmb_{L}(\tilde{x}; \bmx^{(L+1)}) \in \opC^{L-1}$ the vector whose entry on the $b$th row ($b = 1,2,\dots,L-1$) is  $R_{b,L}\tilde{x} + \sum_{\ell = L+1}^B R_{b,\ell} x_\ell$. With these definitions, we find a lower bound on $n_L^\text{future}(x_L;\bmx^{(L+1)})$ using the eigenbound technique in~\cite[Sec.~VII]{stojnic08a}.~Specifically,
\begin{align*}
n_L^\text{future}\big(\tilde{x};\bmx^{(L+1)}\big) 
&\ge \lambda_{L-1}^\text{min}(\bR_{L-1}^H\bR_{L-1}) \\
& \quad \times \min_{\bar\bmx \in \setX^{L-1}}\opnorm{\bar\vecx - \bR_{L-1}^{-1}\bmb_{L}(\tilde{x};\bmx^{(L+1)})}_2^2.
\end{align*}
Here, $\lambda_{L-1}^\text{min}(\bR_{L-1}^H\bR_{L-1})$ is the smallest eigenvalue of the Gram matrix $\bR_{L-1}^H\bR_{L-1}$.
The vector $\bar\vecx \in \setX^{L-1}$ that achieves the minimum is readily obtained by quantizing $\bR_{L-1}^{-1}\bmb_{L}(\tilde{x};\bmx^{(L+1)})$ to the nearest vector in the set $\setX^{L-1}$.


\subsection{Trick 5: Preprune the Search Tree} \label{sec:preprune}

For $\setX = \{\mathsf{x}_1, \mathsf{x}_2, \mathsf{x}_3, \mathsf{x}_4 \}$ as in~\fref{sec:BnB}, large parts of the search tree are redundant since, by symmetry, it holds that $-\mathsf{x}_1 = \mathsf{x}_3$ and that $-\mathsf{x}_2 = \mathsf{x}_4$. Therefore, we can preprune the search tree by excluding symmetric solutions without sacrificing optimality. As illustrated in~\fref{fig:tree}, we exclude all branches (and corresponding subtrees) stemming from $x_B \in \big\{\mathsf{x}_3, \mathsf{x}_4 \big\}$.

\begin{figure}
	\centering
	\includegraphics[width=.95\columnwidth]{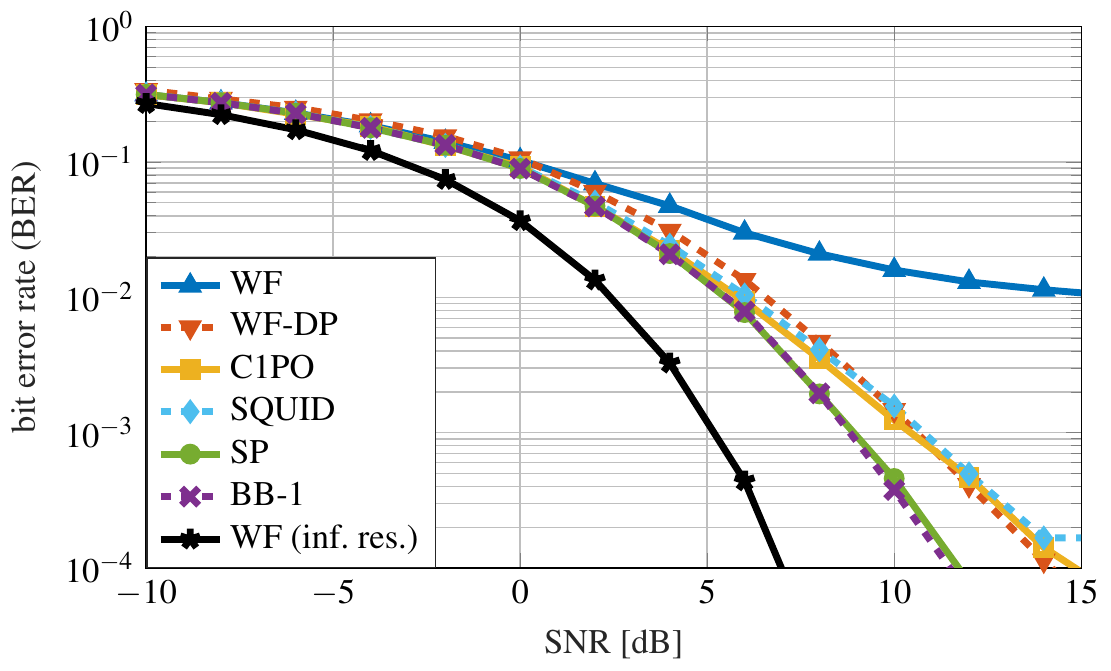}
	\caption{Uncoded BER with QPSK; $B=12$ and $U=3$. Low uncoded BERs can be achieved with the BB-1 precoder and by recently developed state-of-the-art 1-bit precoders.}
	\label{fig:uncoded_ber}
\end{figure}

\section{Simulation Results}

We now investigate the bit error rate (BER) and the complexity of the proposed BB-1 precoder.
For the sake of brevity, we focus on a limited set of system parameters.\footnote{To explore other system configurations, our simulation framework is available on GitHub (https://github.com/quantizedmassivemimo/1bit\_precoding).}
Specifically, we use $B=12$ BS antennas and $U = 3$ users. We consider Rayleigh fading, i.e., the entries of $\bH$ are i.i.d.\ complex circularly symmetric Gaussian distributed with unit variance.

\subsection{BER Performance}

In~\fref{fig:uncoded_ber}, we plot the uncoded BER with QPSK for~{BB-1} as a function of the SNR. For comparison, we also evaluate the BER of state-of-the-art 1-bit precoders.
Specifically, we consider WF precoding, WF precoding with direct perturbation (WF-DP)~\cite[Sec.~3.1]{swindlehurst17a}, convex 1-bit precoding (C1PO)~\cite[Sec.~III-C]{castaneda17a}, squared-infinity norm Douglas-Rachford splitting (SQUID)~\cite[Sec.~IV-B]{jacobsson17d}, and sphere precoding (SP)~\cite[Sec.~IV-C]{jacobsson17d}. 
We also show the BER with WF precoding for the infinite-resolution (no quantization) case. 
First of all, we note that low uncoded BER can be achieved with BB-1. Indeed, the gap to infinite-resolution performance is only $4$\,dB for a target BER of $10^{-3}$. We further note that several of the state-of-the-art precoders, which can be implemented at low computational complexity~(see, e.g,~\cite[Sec.~VI-D]{castaneda17a}), perform close to the BB-1 precoder for low-to-moderate SNR values.

\begin{figure}[!t]
	\centering
	\includegraphics[width=.95\columnwidth]{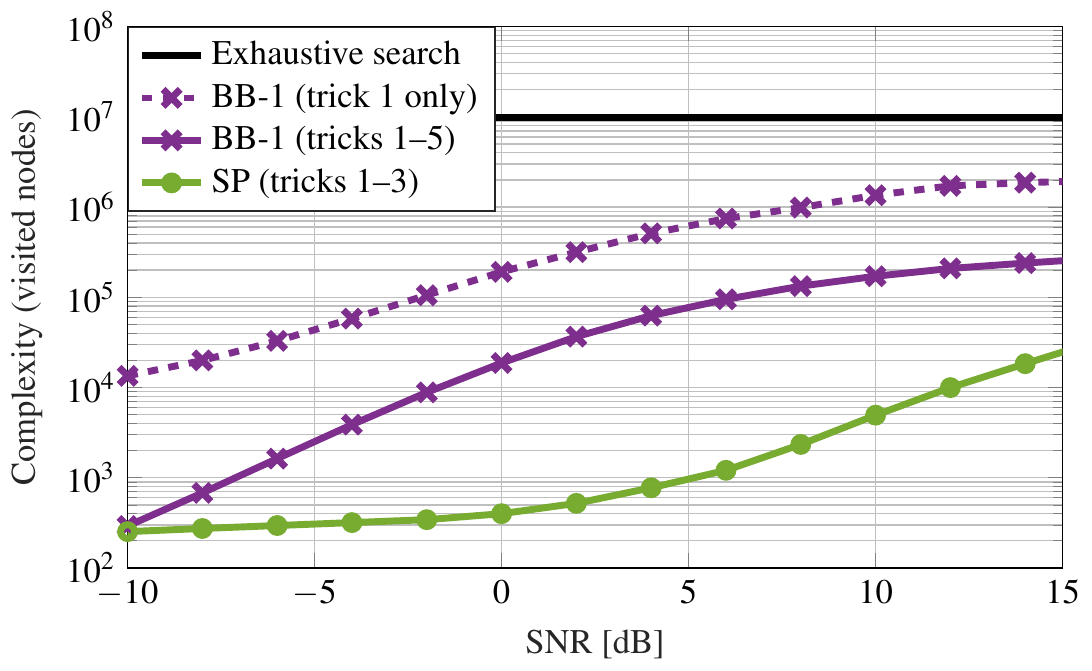}
	\vspace{-0.08cm}
	\caption{Complexity of BB-1 with and without the five tricks proposed in~\fref{sec:tricks}; QPSK, $B=12$, and $U=3$.}
	\label{fig:complexity}
\end{figure}

\subsection{Complexity Impact of the Five Tricks}

In~\fref{fig:complexity}, we show the complexity (measured in terms of the number of nodes visited during a tree search) as a function of the SNR, with and without the tricks presented in~\fref{sec:tricks}. 
We also show the complexity for exhaustive search, for which $\frac{7}{3} \cdot 4^{B-1} - \frac{4}{3}$ nodes are visited during a tree search, and for SP.  
We note that by traversing the tree as in~\fref{sec:traversal}, BB-1 has to visit orders-of-magnitude fewer nodes compared to an exhaustive search, especially at moderate to low SNR values. Indeed, if $N_0$ is small, then the augmented channel matrix $\widetilde\matH$ is ill-conditioned and many eigenvalues of the  Gram matrix are small, resulting in poor pruning behavior. 

By using the tricks proposed in~\fref{sec:radius_init}--\fref{sec:preprune}, the complexity of BB-1 is further reduced drastically.
Note that the tricks in Sections \ref{sec:traversal} to~\ref{sec:sorted} can be used also for SP. Further note that the complexity of SP, which delivers near-optimal performance (cf.~\fref{fig:uncoded_ber}), is noticeably lower than that of~BB-1.

\section{Conclusions and Uses of BB-1}

We have shown how the sum MSE-optimal 1-bit precoding problem can be transformed into a tree search, which is solved exactly for small-to-moderate sized MU-MIMO systems via branch and bound. The resulting BB-1 precoder can be used as a benchmark for other precoding algorithms.
Note that while we focused on 1-bit precoding and QPSK, the proposed branch-and-bound procedure can be applied for any CM transmit alphabet and for any constellation (e.g., 16-QAM). 

We have also shown how the complexity of the branch-and-bound procedure can be significantly reduced by five tricks. To further reduce complexity at the cost of optimality, other tree-traversal strategies (e.g., $K$-best) can be used. 
For millimeter-wave applications where the channel remains constant over multiple symbol intervals, a practical implementation of BB-1 may involve precomputing the precoded vectors for all $\card{\setO}^U$ possible symbol realizations, and storing them in a codebook. 
%

%

%


\bibliographystyle{IEEEtran}
\bibliography{IEEEabrv,confs-jrnls,publishers,svenbib}
\balance

\end{document}